\documentclass[twocolumn]{bmcart}

\usepackage{amsthm,amsmath}
\RequirePackage{hyperref}
\usepackage[utf8]{inputenc} 

\usepackage{comment}
\usepackage{graphicx}

\startlocaldefs
\endlocaldefs

\usepackage{color,soul}
\sethlcolor{yellow}
\renewcommand\hl[1]{#1} 

\begin{document}

\begin{frontmatter}

\begin{fmbox}
\dochead{Research}


\title{Self-adaptive Architectures in IoT Systems: A Systematic Literature Review}


\author[
  addressref={aff1,aff2},
  email={id.alfonso@uniandes.edu.co}
]{\inits{I.A.}\fnm{Iván} \snm{Alfonso}}
\author[
  addressref={aff1},
  email={kj.garces971@uniandes.edu.co}
]{\inits{K.G.}\fnm{Kelly} \snm{Garcés}}
\author[
  addressref={aff1},
  email={hcastro@uniandes.edu.co}
]{\inits{H.C.}\fnm{Harold} \snm{Castro}}
\author[
  addressref={aff2,aff3},
  email={jordi.cabot@icrea.cat}
]{\inits{J.C.}\fnm{Jordi} \snm{Cabot}}


\address[id=aff1]{
  \orgdiv{Department of Systems and Computing Engineering},
  \orgname{Universidad de los Andes},
  \city{Bogotá},
  \cny{Colombia}
}
\address[id=aff2]{
  \orgname{Universitat Oberta de Catalunya},
  \city{Barcelona},
  \cny{Spain}
}
\address[id=aff3]{
  \orgdiv{ICREA},
  \orgname{},
  \city{Barcelona},
  \cny{Spain}
}



\begin{abstractbox}

\begin{abstract} 
Over the past few years, the relevance of the Internet of Things (IoT) has grown significantly and is now a key component of many industrial processes and even a transparent participant in various activities performed in our daily life. IoT systems are subjected to changes in the dynamic environments they operate in. These changes (e.g. variations in bandwidth consumption or new devices joining/leaving) may impact the Quality of Service (QoS) of the IoT system. A number of self-adaptation strategies for IoT architectures to better deal with these changes have been proposed in the literature. Nevertheless, they focus on isolated types of changes. We lack a comprehensive view of the trade-offs of each proposal and how they could be combined to cope \hl{with simultaneous events of different types.}

In this paper, we identify, analyze, and interpret relevant studies related to IoT adaptation and develop a comprehensive and holistic view of the interplay of different dynamic events, \hl{their consequences on QoS}, and the alternatives for the adaptation. To do so, we have conducted a systematic literature review of existing scientific proposals and defined a research agenda for the near future based on the findings and weaknesses identified in the literature.
\end{abstract}



\end{abstractbox}
\end{fmbox}

\end{frontmatter}



\section{Introduction}
\label{sec:intro}

The Internet of Things (IoT) represents a global environment that interconnects the internet with a large number of (cyber-)physical objects such as sensors, vehicles, cell phones, household appliances, cameras, and machines \cite{gubbi2013internet}. IoT aims to facilitate communication and exchange of information to enable new forms of interaction between things and people \cite{cirani2014scalable}. The IoT has transformed and improved the activities we carry out on a daily basis in various aspects such as transport, agriculture, healthcare, industrial automation, and emergency response.

In most IoT systems, it is critical to guarantee the quality of service (QoS) to the users, according to the requirements of the application domain. For example, in continuous monitoring systems, a decrease in the quality could generate wrong or late alerts stemming from the monitored system (imagine the effects of late alerts in monitoring systems in hospitals). Several metrics to measure the quality of service of IoT systems have been proposed. Singh and Baranwal \cite{singh2018quality} classify these into three groups: (1) QoS of communication to measure the quality of network services with metrics such as jitter, bandwidth, performance and efficiency, and network connection time; (2) QoS of things with metrics such as availability, reliability, response time, and security; and (3) QoS of computation to measure computational performance with metrics such as scalability, dynamic availability, and response time.

Satisfying these commitments is challenging due to the dynamic nature of the environment surrounding the IoT system. All types of unexpected events (such as unstable signal strength, growth in the number of connected devices, and software and hardware aging \cite{buyya2019fog, patel2017using}) can happen at any time, posing a risk to the QoS. \hl{This unpredictable behavior of the system can be addressed at runtime by self-adaptive systems. A self-adaptive system modifies its behavior at runtime in response to changes in the system or its environment to ensure a certain quality level} \cite{de2013software}. 

\hl{Self-adaptive systems have been studied for several decades. Wong et. al.} \cite{wong2021self} \hl{classify the evolution of self-adaptive systems into stages. In the first stage (1990-2002), the authors proposed a theoretical model of self-adaptive systems} \cite{cheng1992self, beauquier1999new}. \hl{The second stage (2003-2005) was dominated by studies proposing novel perspectives but without concrete implementations. In particular, Rainbow} \cite{garlan2004rainbow} \hl{was released as one of the foundations of self-adaptive systems. Rainbow, used as a standard, is a framework that involves principles of self-configuration, self-optimization, self-healing, and self-protection in computer systems. In the third stage (2006-2010), research was focused on autonomous and self-adaptive web services. Runtime solutions predominated over design time solutions. For example, in modesl@runtime} \cite{blair2009models} \hl{(the use of software models for adaptive mechanisms to manage complexity in runtime environments) was introduced. The last stage (2011-2020) shows a transition between the domains of research interest. The adaptability of IoT systems and Infrastructure as a Code (IaaS) become the focus. However, the exponential increase and variability of IoT devices, and the unpredictable behavior of the environment introduces self-adaptation challenges to maintain quality levels.} 

This evolution show an increase in the complexity of \hl{IoT architectures.} Traditional IoT systems consisted of two main layers: (1) the device layer, or physical layer, composed by devices (sensors, actuators, and network devices) that generate and send data to the cloud for processing; and (2) the cloud layer made up of servers that store the application logic and process the data generated in the physical layer. 
This architecture has been used for several years. An advantage of such an architecture is that it reduces maintenance costs and application development efforts \cite{patel2017using}. However, new requirements are pushing changes in the design of IoT systems. Centralized cloud-based architectures cannot properly support the constraints of certain IoT applications. The two main reasons for this are: (1) bandwidth limitations and excessive data transmission costs from system devices to the cloud; and (2) communication delay between devices and the cloud, mainly in applications that require real-time data analysis \cite{jiang2017challenges}. These limitations prevent the development of IoT systems that require low-latency responses for large volumes of data to be processed.

\begin{figure}[ht]
  \includegraphics[width=3in]{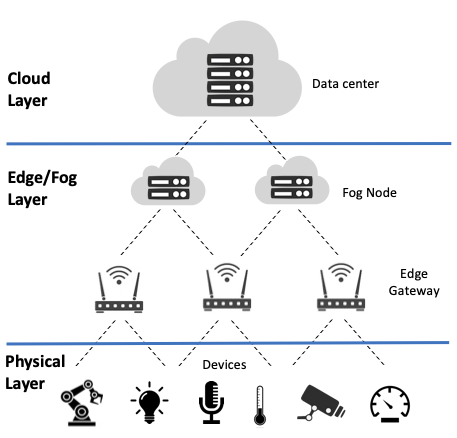}
  \caption{Decentralized architecture for iot systems}
  \label{fig:decentralized-IoT}
\end{figure}

Recently, fog and edge computing have emerged as architectures to address some of the challenges posed by the centralized architecture of cloud-based IoT systems. These architectures are implemented as a layer called edge/fog in the system architecture (Figure \ref{fig:decentralized-IoT}) between the physical layer and the cloud layer. Fog computing is performed on the system's fog nodes while edge computing is performed on edge devices, e.g. gateways. \hl{Nevertheless, both fog and edge computing have a similar purpose: to reduce data travel, latency, and bandwidth consumption, moving computation, storage, communication, control, and decision making closer to the network edge where data is being generated; i.e., in the physical layer} \cite{openfog2017openfog}. Many authors claim that edge and fog are the same, while other authors indicate that fog is a part of edge computing \cite{satyanarayanan2017emergence}. Nevertheless, the combination of edge, fog, and cloud computing makes it possible to build a distributed architecture to guarantee QoS compliance in IoT applications. Fog and edge computing offer advantages mainly in terms of latency and bandwidth usage, since they allow data processing at the edge rather than in the cloud.

Despite the advantages of this more flexible architecture, QoS is still impaired by the dynamic events mentioned above. We will use an illustrating example to give a better understanding of the dynamic events and implications: an IoT system for monitoring and controlling ventilation in underground mines, consisting of hundreds or thousands of sensors and actuators that monitor the mining atmosphere, primarily to ensure the safety of workers. The system monitors the location of workers within the mine, and monitors physical variables such as the concentration of toxic and explosive gases, temperature, and oxygen. It also generates alerts and controls mine ventilation with actuators such as fans and alarms to prevent accidents like poisoning or explosions. However, in daily operation, this system could present problems due to the dynamic environment in the following cases: 

\begin{itemize}
\item The frequency of monitoring toxic and explosive gases may change at certain times, depending on the amount of work and people in the mine. Sensors increase the frequency of monitoring and sending data in the areas of the mine where activity is recorded, which increases bandwidth consumption. On the other hand, when the system detects that the concentration of a hazardous gas exceeds the permitted thresholds, the sensing devices increase their monitoring frequency in the area or throughout the mine. Bandwidth consumption can increase significantly, causing system failures that impact QoS.
\item Some devices in the physical layer of the system (sensors and actuators) change their location within the mine. For example, a sensor can be moved from an abandoned area to work areas where new excavations are made in the mine. However, the system should have the ability to configure the network semi-automatically and provide resources to ensure the mobility of system devices.
\item Finally, wireless devices suffer from aging software \cite{wen2017fog} that sometimes induces problems in the system. For example, it is common to update service or application software to improve security, solve bugs, or improve application performance. Updating device software is not a trivial task when it comes to an IoT system with hundreds or thousands of devices. In addition, applications and services running on the edge/fog and cloud layers also need to be updated by developers.
\end{itemize}

These are just a few of the problems that can be caused by the dynamic environment of an IoT system in the mining domain. However, for other domains, there may be more dynamic environmental events that impact QoS even for systems with distributed architectures. In later sections of this paper, we will discuss dynamic environmental events in other domains. 

Most studies found in the literature individually address particular dynamic events in IoT systems and propose specific strategies to ensure QoS. However, each proposal provides only a partial view (and solution) to the self-adaptation problem. It is necessary to have a comprehensive view of all the different kinds of events (for example, environmental) addressed in the literature. Indeed, we need: (1) a classification of the dynamic events that impact the QoS; (2) a classification of the self-adaptation strategies of the IoT system architecture; and (3) some gaps and challenges in the proposed strategies and their relationships.

In this sense, this paper aims to provide such comprehensive overview of the current state of the art in IoT adaptation. To do so, we conducted a systematic literature review of the proposals in this domain.

The remainder of the paper is organized as follows: 
Section \ref{Sec:method} presents the literature review method, including the research questions, search and selection process, inclusion and exclusion criteria, quality assessment, data extraction, and data analysis. 
Sections \ref{Sec:RQ1} and \ref{Sec:RQ2} address research questions and limitations of current research. Section \ref{Sec:threats-validity} presents threats to the validity of our literature review work.
We present the summary and future directions opportunities in Section \ref{Sec:summary-future-directions}.
In Section \ref{Sec:rel-work}, we analyze related work.
Finally, Section \ref{Sec:Conclusion} presents the conclusions of the review.


\section{Method}
\label{Sec:method}
\begin{figure}
\centering
\includegraphics[width=2.8in]{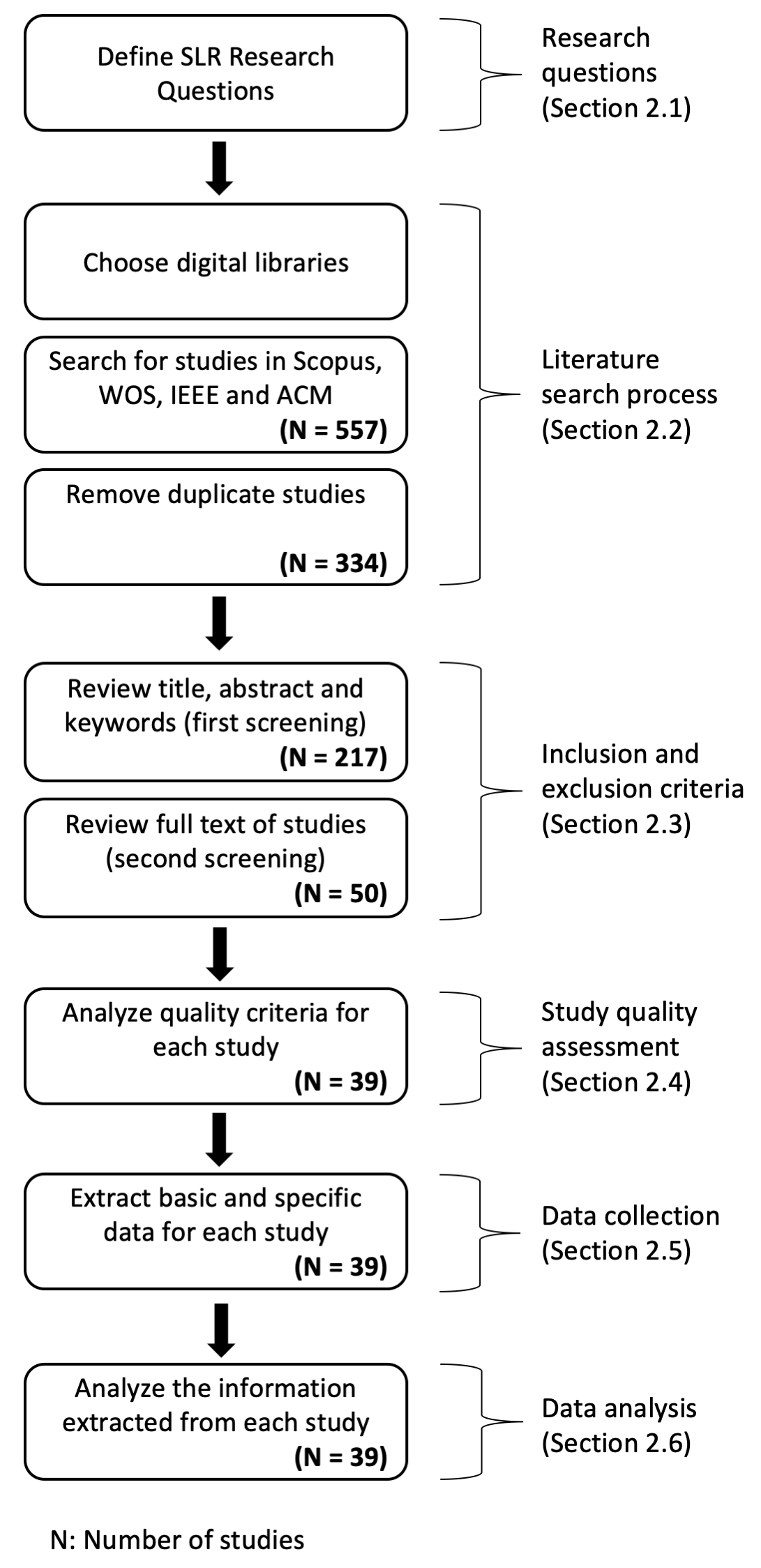}
\caption{Summary of the SLR protocol}
\label{fig:SLRProcess}
\end{figure}

A systematic literature review (SLR) is a methodology used for the identification, analysis, and interpretation of relevant studies to address specific research questions \cite{keele2007guidelines}. Given the complexity of the IoT domain, we decided that an SLR was the best way to systematically reach a comprehensive and fair assessment of this topic. Our SLR consists of six main steps and is based on the methodology proposed by Kitcheham et al. \cite{kitchenham2009systematic}. The steps followed for this SLR are illustrated in Figure \ref{fig:SLRProcess} and documented below.

\subsection{Research questions}
\label{Sec:research-questions}
Our goal is to identify the dynamic environmental events in the physical and edge/fog layers of an IoT system that could impact its QoS and therefore require the trigger of self-adaptations of the system. In addition, we classify the strategies to achieve this self-adaptation. For this purpose, our SLR addresses the following two research questions:
\begin{itemize}
\item RQ1. Which dynamic events present in the edge/fog and physical layers are the main causes for triggering adaptations in an IoT system?
\item RQ2. How do existing solutions adapt their internal behavior and architecture in response to dynamic environmental events in the edge/fog and physical layers to ensure compliance with its requirements?
\end{itemize}

\subsection{Literature search process}
The search process step had three phases \cite{kitchenham2009systematic}: first, we selected the digital libraries; next, we defined the search queries; and finally, we carried out the search and discarded the repeated studies. This section details these phases.

\subsubsection{Digital libraries}
We chose four digital libraries for our search: Scopus, Web of Science (WOS), IEEE Explore, and ACM. These libraries are frequently updated and contain a large number of studies in the area of this research.
\subsubsection{Search Queries}
As shown in Table \ref{table:queries}, we defined \hl{three} search queries. \hl{Keywords such as \textit{IoT} and \textit{cyber-physical} were used to focus the search towards these types of systems; \textit{adapt} or variations of this word (e.g., \textit{adaptation}) and \textit{self-adapt} are keywords to identify adaptation strategies (RQ2) and the events that trigger them (RQ1); \textit{architecture} and \textit{dynamic} keywords to identify studies focused on adaptations on system architecture; \textit{fog}, \textit{edge}, and \textit{osmotic} keywords to retrieve studies using distributed architectures with fog, edge, and osmotic computing (a paradigm to support the efficient execution of IoT services and applications at the network edge} \cite{villari2016osmotic}); and \textit{orchestration} or \textit{choreography} keywords, two resource management techniques at the fog layer of a system. We searched for matches of the aforementioned words in the title, abstract and keywords of the articles.

\begin{table}[ht]
  \caption{Search queries}
  \label{table:queries}
  \begin{tabular}{l l}
  \hline
   & Search query \\
  \hline
  SQ1 & ("fog" OR "edge" OR "osmotic") AND ("IoT" OR \\
   & "internet of things" OR "cyber-physical") AND \\
   & ("architecture") AND ("adapt*" OR "self-adapt*") \\
  SQ2 & "fog" AND "adapt*" AND "architecture" AND \\
   & "orchestration" \\
  SQ3 & ("orchestration" OR "choreography") AND "fog" \\
   & AND "architecture" AND "dynamic" \\
  \hline
  \end{tabular}
\end{table}

\subsubsection{Search Results}
Table \ref{table:StudiesDL} shows the search results; \hl{we obtained 557 studies after applying the three search queries}, out of which 223 were duplicates, for a total of 334 studies.

\begin{table}[ht]
  \caption{Studies per digital library}
  \label{table:StudiesDL}
  \begin{tabular}{l l}
  \hline
  Digital library & Studies found\\
  \hline
  Scopus & 229\\
  Web of Science (WOS) & 120 \\
  IEEE & 176 \\
  ACM & 32 \\
  Total & 557\\
  Total without duplicates & 334\\
  \hline
  \end{tabular}
\end{table}

\subsection{Inclusion and exclusion criteria}
To screen and obtain the primary studies that address the research questions, we defined inclusion and exclusion criteria. We applied two screening phases: in the first screening of the titles, abstracts and keywords, we used three exclusion criteria, to exclude 117 out of the 334 studies. Then, in the second filter we analyzed the full texts, and we discarded 170 additional studies. Finally, using Snowballing to check the list of study citations we included three additional studies, for a total of 50 studies (see Figure \ref{fig:SLRProcess}). The inclusion and exclusion criteria for each screening phase are presented below.

First screening:
\begin{itemize}
\item (Exclusion) \hl{It is not a primary study.} Literature reviews are discarded.
\item (Exclusion) \hl{It is not a journal, conference or workshop paper.}
\item (Exclusion) \hl{The paper} is written in a language other than English
\end{itemize}

Second screening:
\begin{itemize}
\item (Inclusion) The study addresses a dynamic event in IoT systems that impacts QoS.
\item (Inclusion) The study proposes, takes advantage or analyzes a strategy of self-adaptation of architecture for IoT systems.
\end{itemize}

\subsection{Quality assessment}
The quality assessment step consists of reading the studies in detail, and answering the assessment questions to get a quality score for each study. We have defined 5 quality assessment questions as follows:

\begin{itemize}
\item QA1. Are the aims clearly stated? (Yes) the purpose and objectives of the research are clear; (Partly) the aims of the research are stated, but they are not clear; (No) The aims of the research are not stated, and these are not clearer to identify.
\item QA2. Is the research compared to related work? (Yes) the related work is presented and compared to the proposed research; (Partly) the related work is presented, but the contribution of the current research is not differentiated; (No) the related work is not presented.
\item QA3. Is there a clear statement of findings and do they have theoretical support? (Yes) the findings are explained clearly and concisely, and are supported by a theoretical foundation; (Partly) the findings are clearly explained, but they lack theoretical support; (No) findings are not clear and have no foundation or theoretical support.
\item QA4. Do the researchers explain future implications? (Yes) the author presents future work; (No) future work is not presented.
\item QA5. Has the proposed solution been tested in real scenarios? (Yes) The solution is tested in a real scenario; (Partly) the solution is tested in a particular test bed; (No) the solution is not tested in any scenario.
\end{itemize}

The score given to each answer was: Yes = 1, Partly = 0.5, and No = 0. We calculated the quality score for each study and excluded those that scored less than 3, to select the primary studies that would be used for data extraction and analysis. We analyzed 50 studies and excluded eleven because they obtained a quality score of less than three. In total, we have obtained 39 primary studies for the remaining steps of this SLR, and the quality scores for each is presented in Table \ref{table:studies}.

\subsection{Data collection}
\label{sec:data-collection}
The extracted information was stored in an Excel spreadsheet. Table \ref{table:data-collection} shows the Data Collected (DC) for each study and the research question addressed. First, we extracted standard information such as title, authors, and year of publication (DC1 to DC4). Second, we extracted relevant information to address the research questions defined in section \ref{Sec:research-questions}. DC5 records the environmental event addressed by the study, and this information is used to address research question RQ1. DC6 to DC10 are data collected about proposed solutions and strategies to achieve self-adaptations in the IoT system, and this information is used to address research question RQ2.

\begin{table}[ht]
  \caption{Data collection}
  \label{table:data-collection}
    \begin{tabular}{l l l}
    \hline
    \# & Field & RQ\\
    \hline
    DC1 & Author & N/A \\
    DC2 & Title & N/A \\
    DC3 & Year & N/A \\
    DC4 & Publication venue & N/A \\
    DC5 &  Environmental event addressed by the solution & RQ1 \\
    DC6 & Favored quality attributes & RQ2 \\
    DC7 & Adaptation strategies and techniques & RQ2 \\
    DC8 & Architecture description & RQ2 \\
    DC9 & Architectural styles and patterns & RQ2 \\
    DC10 & Key responsibilities of architectural components & RQ2 \\
    \hline
    \end{tabular}
\end{table}

\subsection{Data analysis}
\label{sec:data-analysis}
Table \ref{table:studies} presents the list of the 39 studies relevant to this SLR, with the following information: the assigned identification number (ID), the author, the type of publication, the year of publication, the answers to the quality questions, and the quality score obtained. In the following sections, we will refer to primary studies by the assigned ID code.

\begin{table*}[ht]
\caption{Studies}
\label{table:studies}
        \begin{tabular}{l l l l l l l l l l}
            \hline
            ID & Author & Type & Year & QA1 & QA2 & QA3 & QA4 & QA5 & QA Score \\
            \hline
            S1 & Young, R. et al. \cite{young2018governance} & Conference & 2018 & Y & Y & Y & Y & P & 4.5 \\
            S2 & Wang, J. et al. \cite{wang2017elastic} & Workshop & 2017 & Y & Y & Y & N & P & 3.5 \\
            S3 & Muñoz, R. et al. \cite{munoz2018integration} & Article & 2018 & Y & Y & Y & N & P & 3.5 \\
            S4 & Cheng, B. et al. \cite{cheng2015geelytics} & Conference & 2015 & Y & Y & Y & Y & N & 4 \\
            S5 & Kimovski, D. et al. \cite{kimovski2018adaptive} & Conference & 2018 & Y & Y & Y & Y & P & 4.5 \\
            S6 & Young, R. et al. \cite{young2018dynamic} & Conference & 2018 & Y & Y & Y & Y & P & 4.5 \\
            S7 & Tseng, C. et al. \cite{tseng2018extending} & Conference & 2018 & Y & N & Y & Y & P & 3.5 \\
            S8 & Peros, S. et al. \cite{peros2018dynamic} & Conference & 2018 & Y & Y & Y & Y & P & 4.5 \\
            S9 & Rausch, T. et al. \cite{rausch2018emma} & Conference & 2018 & Y & Y & Y & N & Y & 4 \\
            S10 & Pahl, C. et al. \cite{pahl2018architecture} & Conference & 2018 & Y & Y & Y & N & P & 3.5 \\
            S11 & Lorenzo, B. et al. \cite{lorenzo2018robust} & Article & 2018 & Y & Y & Y & Y & N & 4 \\
            S12 & Prabavathy, S. et al. \cite{prabavathy2018design} & Article & 2018 & Y & Y & Y & Y & P & 4.5 \\
            S13 & Yigitoglu, E. et al. \cite{yigitoglu2017foggy} & Conference & 2017 & Y & Y & Y & Y & P & 4.5 \\
            S14 & Morabito, R. et al. \cite{morabito2017framework} & Workshop & 2017 & P & P & Y & Y & P & 3.5 \\
            S15 & Desikan, K. S. et al. \cite{desikan2017novel} & Workshop & 2017 & Y & Y & Y & N & P & 3.5 \\
            S16 & de Brito, M. S. et al. \cite{de2017service} & Conference & 2017 & Y & Y & Y & Y & N & 4 \\
            S17 & Velasquez, K. et al. \cite{velasquez2017service}& Conference & 2017 & Y & P & Y & Y & P & 4 \\
            S18 & Flores, H. et al. \cite{flores2017large} & Conference & 2017 & Y & N & Y & Y & P & 3.5 \\
            S19 & Pizzolli, D. et al. \cite{pizzolli2016cloud4iot} & Conference & 2016 & Y & N & P & Y & P & 3 \\
            S20 & Montero, D. et al. \cite{montero2016offloading} & Conference & 2016 & Y & Y & Y & Y & P & 4.5 \\
            S21 & Chen, L. et al. \cite{chen2018adaptive} & Article & 2018 & Y & Y & Y & Y & Y & 5 \\
            S22 & Mass, J. et al. \cite{mass2018context} & Conference & 2018 & Y & Y & Y & Y & Y & 5 \\
            S23 & Li, X. et al. \cite{li2018adaptive} & Article & 2018 & Y & Y & Y & N & Y & 4 \\
            S24 & Suganuma, T. et al. \cite{suganuma2018multiagent} & Article & 2018 & Y & Y & Y & Y & Y & 5 \\
            S25 & Deng, G. et al. \cite{deng2018application} & Conference & 2018 & Y & Y & Y & N & Y & 4 \\
            S26 & Sami, H. et al. \cite{sami2018towards} & Conference & 2018 & Y & Y & Y & Y & Y & 5 \\
            S27 & Wu, D. et al. \cite{wu2019fog} & Conference & 2019 & Y & Y & Y & N & Y & 4 \\
            S28 & Skarlat, O. et al. \cite{skarlat2018framework} & Conference & 2019 & Y & Y & Y & Y & Y & 5 \\
            S29 & Mechalikh, C. et al. \cite{mechalikh2019scalable} & Conference & 2019 & Y & Y & Y & Y & Y & 5 \\
            S30 & Castillo, E. et al. \cite{castillo2019iot} & Conference & 2019 & Y & P & Y & N & Y & 3.5 \\
            S31 & Breitbach, M. et al. \cite{breitbach2019context} & Conference & 2019 & Y & Y & Y & Y & Y & 5 \\
            S32 & Torres Neto, J. et al. \cite{torres2019exploiting} & Article & 2019 & Y & Y & Y & Y & Y & 5 \\
            S33 & Theodorou, V. et al. \cite{theodorou2019glt} & Workshop & 2019 & Y & N & Y & Y & P & 3.5 \\
            S34 & Guntha, R. \cite{guntha2019iot} & Conference & 2019 & Y & Y & Y & N & P & 3.5 \\
            S35 & Jutila, M. \cite{jutila2016adaptive} & Article & 2016 & Y & Y & Y & Y & Y & 5 \\
            S36 & Cui, K. et al. \cite{cui2019joint} & Conference & 2019 & Y & Y & Y & Y & Y & 5 \\
            S37 & Bedhief, I. et al. \cite{bedhief2019toward} & Conference & 2019 & Y & Y & Y & P & N & 3.5 \\
            S38 & Asif-Ur-Rahman, Md et al. \cite{asif2018toward} & Article & 2019 & Y & Y & Y & Y & Y & 5 \\
            S39 & Yousefpour, A. et al. \cite{yousefpour2019fogplan} & Article & 2019 & Y & Y & Y & Y & Y & 5 \\
            \hline
        \end{tabular}
\end{table*}

\begin{figure}
  \includegraphics[width=3in]{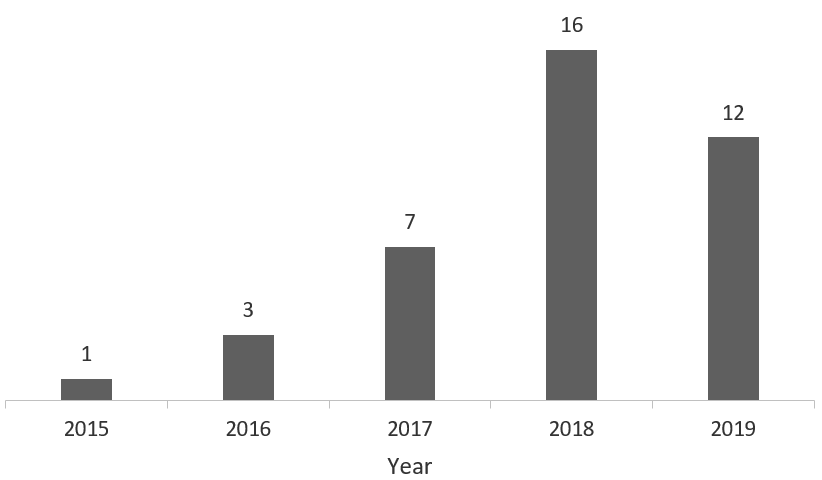}
  \caption{Number of studies published per year}
  \label{fig:Studies-per-year}
\end{figure}

From the standard information extracted from the papers, we can note that the relevant publications for this SLR are relatively recent. The largest number of studies were published in recent years: 12 studies from 2019, 16 studies from 2018, 7 studies from 2017, 3 studies from 2016, and one study from 2015 (see figure \ref{fig:Studies-per-year}). \hl{As for the type of publication}, 25 are conference publications, 10 are journal publications, and 4 are workshop publications.

\section{RQ1: Which dynamic events present in the edge/fog and physical layers are the main causes for triggering adaptations in an IoT system?}
\label{Sec:RQ1}

In this section, we present the information and results obtained from the analysis of the data extracted to address research question RQ1. The dynamic events that force an adaptation in the IoT system are presented and discussed in Section \ref{subsection:dynamic-events}. Finally, we discuss monitored QoS metrics for detecting dynamic events in Section \ref{sub:metrics-monitored}.

\subsection{Dynamic events}
\label{subsection:dynamic-events}
Table \ref{table:dynamic-events} provides an overview of the answer to research question RQ1. The table presents a classification of the dynamic environmental events present in the edge/fog and physical layers and the studies that addressed that event. We propose this list of events that we have obtained from the detailed analysis of the studies. We then classify each study according to the event it addresses. \hl{Strategies for adapting the system in response to these events are discussed in Section} \ref{Sec:RQ2}.

\begin{table*}[ht]
  \caption{Dynamic environmental events}
  \label{table:dynamic-events}
  \begin{tabular}{l l l l}
  \hline
  ID & Dynamic event & Studies \\
  \hline
    E1 & \hl{Client mobility} & S5, S9, S10, S17, S19, S20, S22, S29\\ 
    E2 & Dynamic data transfer rate & S3, S6, S7, S11, S15, S18, S19, S21, S26, S32, S39\\
    E3 & Important event detected by sensors & S1, S2, S8, S24, S27, S31, S36, S38\\ 
    E4 & Failures and software aging & S4, S13, S14, S16, S28\\ 
    E5 & Network connectivity & S1, S23, S25, S30, S33, S34, S35, S37\\ 
    E6 & Cyber-attacks in IoT applications & S12 \\ 
  \hline
  \end{tabular}
\end{table*}

\subsubsection{Client mobility}
Mobile devices such as cell phones or automobiles produce events in the physical layer of the IoT system, causing challenges to ensure QoS. When the system devices change their location, it is necessary to make network reconfigurations, storage synchronizations, and rescheduling of processes among the edge/fog nodes based on available resources. For example, \hl{client mobility} is one of the dynamic events present in the example illustrated in Section \ref{sec:intro} because it is necessary to move sensors through the tunnels and work areas in the mine.

Eight studies included in this SLR (S5, S9, S10, S17, S19, S20, S22, and S29) address the mobility of clients in the IoT system, which, through different techniques, seek to provide the resources and services at the edge/fog layer to efficiently manage mobility. For example, in S19 a case study of client mobility is addressed. This case consists of patients wearing a device (without 3G/4G connection) to monitor health parameters such as temperature and heart pressure. When the patient leaves his/her home and moves to the hospital, a gateway in the hospital automatically discovers the wearable device and assigns or associates a monitoring service to it.

Client mobility is an event or requirement of IoT systems that poses challenges due to the constant movement of devices, the heterogeneity of communication technologies, and resources, which can be requested on demand simultaneously by multiple devices in different locations \cite{santos2018fog}. When a device changes location, \hl{a series of} steps are performed: (1) this change has to be automatically detected; (2) the availability of resources must be guaranteed to deploy the service in the edge/fog nodes in order to manage that device; and (3) in case the device changes location again, it is evaluated if it must be connected to other edge/fog nodes that are closer to obtain better latency.

Monitoring and detecting client mobility depend on the communication protocol between the client (devices) and the edge/fog layer nodes. In scenarios with devices that use low level communication protocols (e.g. Bluetooth and WiFi), it is possible to discover the mobile device in a coverage area and automatically associate it to an IoT gateway, as suggested in S19. 

S5 proposes a method of \hl{client mobility discovery using MQTT, a standard messaging protocol. MQTT is frequently used in IoT systems due} to the advantages of the publication/subscription pattern regarding scalability, asynchronism and decoupling between clients. MQTT architecture uses one or several nodes called Brokers to manage the network, to receive messages from publishers, and to send messages to subscribers. S9 proposes a distributed QoS-aware MQTT middleware for edge computing, addressing the mobility of clients. When new clients join the system network, a controller searches and maps the broker that offers less latency to the client. In S9, the mobility of clients is detected through the brokers. Every time a client joins the IoT system, it subscribes to one of the MQTT broker's topics. The broker has the capacity to monitor different factors, such as the number of subscribed clients, and detect when a new client subscribes.

\subsubsection{Dynamic data transfer rate}
The data transmission rate of the devices is another dynamic event that significantly influences the system's QoS. In IoT systems, the data transmission rate from the physical layer to the edge/fog layer may vary depending on the circumstances, objects, or conditions in which the devices are surrounded. The system devices may increase or decrease the frequency of data transmission due to different stimuli. For example, to reduce power consumption in an office building, the sensor devices of the IoT temperature monitoring system are scheduled to raise the data transmission rate during business hours and lower the monitoring frequency and transmission rate during unmanned hours. \textit{Dynamic data transfer rate} event, which is the most discussed topic in the literature, is addressed by eleven studies (S3, S6, S7, S11, S15, S18, S19, S21, S26, S32, and S39) analyzed in this SLR. These studies analyze scenarios with great variability \hl{in the data collected and data transfer rates of IoT devices.}

The consequences of this dynamic event in IoT systems commonly lead to increased latency and unavailability of system services because increased data volume could congest the network and generate bottlenecks. In addition, this dynamic event implies increase in data to be analyzed or processed by the edge devices, which will likely have limited computer resources. Therefore, the edge nodes could be overloaded with processing work resulting in delays, downtimes, or unavailability.

Two monitoring techniques are used to identify changes in the data transfer rate. The first technique is to watch the computational resources consumed by the edge/fog nodes. The \%CPU used is the most commonly used metric by studies (S6, S7, S18, and S32) to identify when an edge/fog node is overloaded due to the increased data it must process. Although the \%CPU does not accurately measure the data transfer rate, it is used to detect an increase or decrease in the amount of data to be processed by the node. Increasing the amount of data that arrives at a node to be attended to or analyzed also increases the processing tasks and the \%CPU used. The second technique to detect data transfer rate variations is to monitor the network, as proposed by S3, S11, S15, S19, S21. For example, in S3, IoT Flow Monitors are deployed to supervise the average bandwidth of the aggregated IoT traffic and to detect IoT-traffic congestion. Wireshark\footnote{\url{https://www.wireshark.org}} is one of the tools used to monitor network IoT traffic.

\subsubsection{Important event detected by sensors}
When an alert or alarm is generated by sensor data in an IoT monitoring system, a set of tasks is triggered to inform the end user and/or control the emergency. These tasks may increase network, processing, and storage consumption at some layer of the system architecture (physical, edge/fog, and cloud). For example, in a smart city when a vehicular accident is detected by video surveillance cameras, the accident is immediately communicated to authorities (e.g., the police) and medical centers. New processing tasks begin to run in edge/fog nodes or cloud servers: 1) there are increases in the processing and storage of video taken by surveillance cameras; 2) visual alerts are generated to other drivers on the road; 3) tasks are executed to synchronize streetlights to address the emergency and reduce vehicle traffic. Studies S1, S2, S8, S24, S27, S31, S36, and S38 detect important events from sensors. For example, in study S1, changes in weather conditions are detected (e.g., when it starts raining). In study S2, emergencies are detected through a video surveillance system. In study S8, events are detected when one of the IoT devices breaks a rule configured by the user (e.g. when a motion sensor is activated). In study S24, emergency situations such as sudden illness of a group of athletes is detected by wearable vital sensors. Studies such as S31 and S36 do not monitor specific events in the physical layer but propose solutions to address these types of events that commonly require the deployment of additional services at the edge and fog nodes.

System tasks generated by alarms or alerts commonly require additional network, processing, or storage resources. Some systems react to these types of events by increasing monitoring frequencies. Other systems react by deploying or running new tasks on the edge/fog nodes. These tasks and system reactions can affect network connectivity due to increased bandwidth consumption and increased processing at the edge/fog nodes.

The events detected by the system sensors depend on the domain of the application and the IoT system. For some systems, it is important to monitor weather conditions because they significantly influence the performance of wireless networks such as Wi-Fi \cite{bri2012weather}. Other systems only need to monitor the processes of the domain itself. These events are detected by analyzing the data coming from the devices of the physical layer of the system, and a component of the system architecture is responsible for analyzing the data to detect the event. This component commonly checks that the data sensors are within an expected range. For example, the approach proposed in S1 focuses on a connected vehicle use case where weather conditions are monitored. The vehicle sends data frequently to a central controller to predict driver alertness. When rain is detected, it is assumed that the quality of the communication between the vehicle and the node decreases. The size of the subset of data sent to the central controller is then reduced, selecting only the critical data for driver alertness prediction.

\subsubsection{Failures and software aging}
The software embedded in the devices, nodes, and servers of an IoT system needs to be updated and redeployed by developers to fix service errors, improve application performance, improve system security, etc. Some upgrades or deployments of system services and application software may involve adaptations to the layers of the system architecture. First, when new services are deployed at edge/fog nodes, it may be necessary to adapt the bindings (e.g., service registry, network topology) established between the services deployed in the nodes and the components that consume said services to ensure the communication. Second, software upgrades are sometimes unsuccessful due to storage, hardware, or connectivity failures. In these cases, the system should detect the problem and fix it. Third, the physical layer and edge/fog devices have limited processing capabilities that may bring risks to successful software upgrades. This implies increased latency and, in some cases, unresponsive services.

S14 proposes a flexible architecture that allows for the network to be dynamically adapted and for the containers to be routed in the fog nodes every time new software deployments could change the operational chain. S28 propose FogFrame, a framework that reacts dynamically to failures or overloads in fog nodes. FogFrame redeploys or redistributes software to reduce node overload or to ensure availability when a fog node fails. S4, S13, and S16 provide frameworks for automating software deployments in fog layer nodes and dynamically adapting the location of containers and the network topology. For example, S13 proposes Foggy, a framework to facilitate and automate the deployment of software in fog nodes of an IoT system. The deployment of applications in fog nodes of the system and the management of resources is handled by an orchestration server. Foggy was based on the use of Docker\footnote{\url{https://www.docker.com}} containers and deployment rules to facilitate dynamic resource provisioning. Container allocation decisions are defined by the developer through deployment rules. For example, the developer can create a rule to deploy a software version to all fog nodes with RAM greater than 2GB. These deployment rules give the developer control over deployment location decisions. The Orchestration Server monitors the use of resources and dynamically adapts the placement of the containers in the nodes. However, Foggy does not detect faults in the containers at runtime, nor does it perform system adaptations such as rollback, redeployment, or movement of software versions.

To detect software failures and aging, it is necessary to constantly monitor the availability of services, nodes, and the status of software containers and/or virtual machines. Two techniques are commonly used to detect these types of failures: (1) ping/echo is an asynchronous request/reply message to determine reachability and the round-trip delay, but this technique is only for nodes interconnected via IP; and (2) heartbeat is a fault detection mechanism that consists of exchanging messages periodically between the node and a monitoring component \cite{bass2003software}. 


\subsubsection{Network connectivity}
According to S3, the main network requirements for IoT services are low latency, high-speed traffic, large capacity traffic, and massive connections. Although these requirements depend on the domain of the IoT application, most systems require the fulfillment of at least one of these.
IoT systems constantly present variations in network connectivity characteristics that make it difficult to meet network requirements. These variations, mainly present in wireless communications, can generate negative effects on the transmission and reception of data between the system's devices, nodes, and cloud servers: (1) out-of-date information due to communication delays; (2) incomplete information due to intermittent or interrupted communication; (3) unavailability of services or system applications due to lost or broken communication. 

Network characteristics may be affected by changes in weather \cite{bri2012weather}, variations in system power voltage, wireless signals that interrupt communication, high bandwidth consumption, or other external factors that are normally ignored. To detect deterioration in the quality of communications between system devices and the edge/fog nodes, it is necessary to monitor the network. For example, in S1, a component is designed to monitor and detect changes in network connectivity. When the monitor detects that the network connectivity is less than 50\%, the amount of data sent by the devices from the physical layer to the edge node is decreased and prioritized. To monitor and detect changes in the quality of communication, it is necessary to monitor network metrics such as latency, bandwidth, and lost packets. In S33, S34, and S35 the communication latency between the devices and the nodes or servers that process the data is monitored. When the latency exceeds a predefined threshold, adaptations of data flow reconfiguration and task offloading are performed. In S34, the state of the connection to the cloud is monitored. If the network connection to the cloud is lost, sensor data analysis tasks hosted in the cloud are offloaded to the fog layer nodes.

\subsubsection{Cyber-attacks in IoT applications}
Although the security topic was not intentionally addressed in this study, we found the work of Prabavathy et al. (S12), which proposes a strategy based on the use of fog computing to detect cyber-attacks. The threats that come from the data of the physical layer devices towards the edge/fog layers and cloud are events induced by attackers that violate the confidentiality, integrity, and availability of the system. In an IoT system, sensors and actuator devices frequently capture and share personal data from our daily life, detect critical physical variables in industrial processes, and control the vehicular flow in a city. The impact of an attack on the devices in any of the layers of the architecture can cause loss of critical information, disasters in the processes that control the system, and unavailability of the system, among others. Therefore, it is essential to ensure the security of the IoT system by designing self-adaptation techniques to defend against attacks.

Bass et al. \cite{bass2003software} propose four techniques for detecting attacks on software systems: (1) \textit{detect intrusion} consists of comparing network traffic patterns with known malicious behavior patterns stored in a database; (2) \textit{detect service denial} consists of comparing network traffic entering a system with historical profiles of known denial of service attacks; (3) \textit{verify message integrity} consists of using checksum or hash values to validate the integrity of message information; and (4) \textit{detect message delay} consists of detecting potential man-in-the-middle attacks. These strategies can be adapted to detect attacks on IoT systems. In particular, S12 detects intrusion by implementing an Extreme Learning Machine (ELM) algorithm on the system's fog nodes. ELM is a fast learning algorithm for a hidden single-layer neural network \cite{huang2015trends}, it is suitable for real-time applications due to the low performance of problem solving. In S12, each data packet that is sent from the physical layer to the fog layer is analyzed by the algorithm. This algorithm can detect attacks from different categories including denial of service, user to root, probe-response, and remote to local.

\subsection{QoS metrics monitored}
\label{sub:metrics-monitored}

Monitoring is an important task to detect dynamic events in the IoT systems. These events are detected by analyzing metrics about node resource consumption (such as CPU, memory, and energy consumption), network behavior (such as bandwidth consumption and communication latency), and availability. Table \ref{table:monitored-metrics} presents the monitored metrics to detect the dynamic events for each study. 
The resource consumption in the edge/fog nodes is the most monitored feature to detect events. In particular, CPU and memory consumption are used to detect three of the dynamic events: \textit{Client mobility}, \textit{Dynamic data transfer}, and \textit{Failures and software aging}. Sensor data (column 9) is not a QoS metric, but its analysis is used to detect the dynamic events \textit{Important event detected by sensors}, \textit{Network connectivity}, and \textit{Cyber-attacks in IoT applications}. Availability and Latency are seldom monitored metrics to detect dynamic events. However, ensuring low latency is one of the important requirements for real-time applications. Similarly, ensuring the availability of services and applications in IoT systems is also a common requirement. S10, S20, S22, and S29 are not included in Table \ref{table:monitored-metrics} because they do not monitor any QoS metrics. These four studies address the dynamic event client mobility, which they detect by identifying new clients joining or leaving the system. Studies S31 and S36 do not focus on the detection of the dynamic event, instead they cover the architectural adaptations to cope with the event. For this reason, these two studies are not included in Table \ref{table:monitored-metrics}.

The QoS metrics monitoring conducted by the studies is carried out in the edge/fog and cloud layers of the system, but not in the physical layer devices. \hl{The World Wide Web Consortium} \cite{W3C} proposes an ontology of non-functional properties for IoT devices including accuracy, sensitivity, response time, drift, and frequency. The monitoring of these properties is not trivial due to the heterogeneity of IoT devices, difference in firmwares, and variability of communication protocols. However, constantly monitoring these properties at the physical layer of the system would allow for improved QoS. For example, it is possible to avoid device failures by scheduling preventive maintenance after analyzing the information on the accuracy and sensitivity of the sensors.

\begin{table*}
\caption{Monitored metrics}
\label{table:monitored-metrics}
        \begin{tabular}{l l l l l l l l l l}
            \hline
            Event & Study & CPU & Memory & Storage & Bandwidth & Availability & Latency & Sensor data \\
            \hline
            E1 & S5 & X & X & & & & & \\
            E1 & S9 & & & & & & X & \\
            E1 & S17 & X & X & X & & & & \\
            E1/E2 & S19 & & & X & & X & \\
            E2 & S3 & & & X & & & \\
            E2 & S6 & X & & & & & & \\
            E2 & S7 & X & & & & & & \\
            E2 & S11 & & & & X & & & \\
            E2 & S15 & & & & & & X & \\
            E2 & S18 & X & X & X & & & & \\
            E2 & S21 & & & & & X & \\
            E2 & S32 & X & & & & & \\
            E3 & S1 & & & & X & & & X \\
            E3 & S2 & & & & & & & X \\
            E3 & S8 & & & & & & & X \\
            E3 & S24 & & & & & & & X \\
            E3 & S27 & & & & & & & X \\
            E3 & S38 & & & & & & & X \\
            E4 & S4 & X & X & X & X & X & X & \\
            E4 & S13 & X & X & X & X & & & \\
            E4 & S14 & X & X & X & & & & \\
            E4 & S16 & X & X & X & & & & \\
            E4 & S28 & X & X & X & & & & \\
            E5 & S1 & & & & X & & & X \\
            E5 & S23 & & & & & & X & \\
            E5 & S25 & & & & X & & X & \\
            E5 & S30 & & & & & & X & \\
            E5 & S33 & & & & & & X & \\
            E5 & S34 & & & & X & & & \\
            E5 & S35 & & & & & & X & \\
            E5 & S37 & & & & X & & & \\
            E5 & S39 & & & & & & X & \\
            E6 & S12 & & & & & & & X \\
            \hline
        \end{tabular}
\end{table*}


\section{RQ2: How do existing solutions adapt their internal behavior and architecture in response to dynamic environmental events in the edge/fog and physical layers to ensure compliance with its requirements?}
\label{Sec:RQ2}

In this section, we address the research question RQ2. In Section \ref{subsection:adaptations}, we discuss the adaptive strategies used by the studies to support the dynamic events outlined in the previous section. We then discuss the relationship between dynamic events and adaptive strategies in Section \ref{sub:rel-events-adaptations}.

\subsection{Adaptive strategies}
\label{subsection:adaptations}
Table \ref{table:adaptations}, which presents a classification of the strategies used by each study to support specific dynamic events, provides a preliminary answer to research question RQ2. Similar to the classification of dynamic events (\ref{subsection:dynamic-events}), we propose this list of adaptations after analyzing the studies in detail. Study S12 is not included in Table \ref{table:adaptations} because it does not address an adaptation strategy. This study monitors, detects, and classifies attacks coming from the physical layer devices of the IoT system, but the system is not adapted to these attacks. Although the topic of this SLR is not the security of the IoT system and study S12 does not perform any adaptation strategy, we include it in this SLR because it is the only one that addresses dynamic event E6 in our list of studies, and it contributes to solve research question RQ1.

The adaptive strategies are described below.

\begin{table*}[ht]
  \caption{Adaptations}
  \label{table:adaptations}
  \begin{tabular}{l l l l}
  \hline
  ID & Adaptation & Studies \\
  \hline
   A1 & Data flow reconfiguration & S3, S5, S8, S9, S10, S11, S14, S15, S17, S19, S20, S23, S25, S35, S37, S38\\
   A2 & Auto scaling of services and applications & S2, S7, S17, S18, S19, S22, S31, S39\\
   A3 & Software deployment and upgrade & S4, S13, S16, S28 \\
   A4 & Offloading tasks & S1, S6, S21, S26, S27, S29, S30, S32, S33, S34, S36\\
  \hline
  \end{tabular}
\end{table*}

\subsubsection{Data flow reconfiguration}
The routing of data traveling from the physical layer to the Edge/Fog or cloud layer is modified mainly to improve latency. The direction of the data flow and the devices involved in communication, such as gateways and messaging servers, are strategically selected to carry the data to the nodes that perform the processing. 

The data flow reconfiguration strategy is used to address three dynamic events: \textit{Client mobility}, \textit{Dynamic data transfer rate}, \textit{Important event detected by sensors}, and \textit{Network connectivity}. Studies addressing client mobility (S5, S9, S10, S17, S19, and S20) use this technique to control communication between physical layer devices and edge/fog and cloud layer nodes. When a physical layer device changes location, it is necessary to identify the edge/fog nodes with the most optimal position to provide the lowest communication latency with the device. Additionally, it is necessary to ensure that these edge/fog nodes have the resources available to support the new device load. 

In S9, an edge-enabled publish-subscribe middleware is proposed to address client mobility challenges. The data flow between clients and brokers, i.e., the servers that implement the MQTT server protocol, is reconfigured to optimize communication latency. When there is client mobility, the least communication latency with the MQTT broker is guaranteed. However, the availability of resources of the edge/fog node that hosts the broker is not monitored. If a group of clients is assigned to a broker that has little processing capacity, the broker could be overloaded and could fail. Additionally, there are no mechanisms to autonomously auto-scale the brokers in the edge nodes and register them in the system. Auto-scaling (\ref{Subsec:auto-scaling}) is another strategy used to address some dynamic events, and this strategy is complemented by data flow reconfiguration.

To adapt the IoT system to cope with the dynamic events \textit{Dynamic data transfer rate}, \textit{Important event detected by sensors}, and \textit{Network connectivity}, some authors propose to reconfigure the data flow with the aim of balancing the load between the edge/fog nodes, or to redirect the data flow to the node with the best conditions (resource availability and lower response latency). For example, S8 proposes a framework that enables the developer to specify dynamic QoS rules. A rule is made up of a source device (e.g. a video camera), a target device (e.g., a web server), a rule activation event (e.g. when a system sensor detects motion), and a QoS requirement that must be guaranteed (e.g. 200ms communication latency between source and target). When the event configured in the rule is triggered, the path of the data flow between the source and the destination is reconfigured to establish the optimal path through a set of switches. \hl{This architecture assumes} that there are several switches that enable communication between the physical layer devices and the cloud layer. However, the edge/fog layer is not included to do edge processing, which could improve system QoS by lowering latency and bandwidth. The system architecture proposed in S8 assumes that the edge/fog layer is composed of devices that only serve the function of relaying the data, but the data processing capacity in the edge devices is ignored. Additionally, it is necessary to consider \hl{using the} MQTT protocol and broker for communication which offers lower power consumption and low latency due to its very small message header and packet message size (approximately 2 bytes) \cite{wukkadada2018comparison}.

The software-defined network (SDN) is a network management technology commonly adopted by studies that propose the strategy of adaptation (\textit{Data flow reconfiguration}). S14, S23, S25, S37, and S38, deploy SDN to flexibly manage network resources according to changing system conditions. The SDN controller has the functionalities to configure the data flows through the system devices. Several algorithms have been proposed to optimize data flows and ensure QoS. For example, S25 proposes a routing algorithm that finds the lowest cost routes based on latency, available bandwidth, and lost data packets. S23 proposes a routing algorithm to optimize data traffic by considering QoS metrics such as latency and power consumption. Dynamic management of network resources and dynamic flow is possible through the SDN controller.

\subsubsection{Auto scaling of services and applications}
\label{Subsec:auto-scaling}

This strategy consists of automatically deploying or terminating replicated services and applications on the system's edge/fog nodes or cloud servers. Auto-scaling is used to ensure stable application performance, and it is one of the most widely used techniques in web applications deployed in the cloud. Auto-scaling is also used in IoT systems but with additional concerns to address. A challenge of auto-scaling at the edge/fog layer is related to the selection of the best node for the deployment and execution of the service or application. While the cloud layer has a large amount of network, processing, and storage resources, the edge/fog layer has limited resources. For this reason, when scaling an application at the edge/fog layer, it is necessary to strategically select the node that has availability of the necessary computing resources and that offers the greatest communication latency benefits with the physical layer devices.

According to Table \ref{table:events-strategies}, auto-scaling of services and applications is a technique used to address three dynamic events in IoT systems: (1) for the \textit{Client mobility} event, the applications are auto-scaled to attend the requests of new customers that connect a cluster of edge/fog nodes; (2) for the \textit{Dynamic data transfer rate} event, it is necessary to auto-scale the applications and services in the edge/fog nodes to support the growth of data that are processed; (3) in some domains, the \textit{Important event detected by sensors} requires auto-scaling of applications to support the growth of monitoring frequencies in cases of system emergency.

In S2, an auto-scaling method is proposed for a distributed intelligent urban surveillance system. The proposed architecture has three layers: video cameras in the physical layer, desktops in the edge layer to analyze the video information, and cloud servers that host the web application for the end user. When the video cameras detect an emergency, the frame rates of video capturing increase and image analysis for some objects turn to high-priority tasks. The system then scales the data analysis application by deploying virtual machines to the edge nodes closest to the emergency site. However, deploying the application at the node closest to the physical layer device does not always guarantee the best performance. Other factors such as network latency and node bandwidth consumption should be considered for application allocation decisions. Additionally, the use of virtual machines has limitations given the resource scarcity that characterizes edge nodes. Other virtualization technologies such as containers have advantages for deploying applications to edge/fog layer nodes. In particular, the reduced size of the images and the low startup time are advantages that make containers suitable for IoT systems.

\subsubsection{Software deployment and upgrade}
The process of deploying and updating software in a semi-automatic way is one of the strategies used to solve problems, correct software issues, improve application performance, and improve system security. 

S4, S13, S16, S28, and S39 perform software deployment and movement of services remotely in the edge/fog nodes of the system. Containerization is one of the most used technologies that facilitates the semi-automatic deployment of software, given the reduced size of images and the low start time compared to virtual machines. These three studies (S4, S13, S28, and S39) use docker technology to package and run the software versions in containers on Fog nodes. S13 proposes Foggy, a framework for continuous automated deployment in fog nodes. Foggy allows for the definition of four software allocation rules in Fog nodes: (1) software deployment on a specific node; (2) software deployment on a specific number of nodes that match a hardware feature (i.e., three nodes that have 2GB of memory); (3) software deployment on a group of nodes that comply with a hardware feature (i.e., all nodes that have 4GB of memory); and (4) software deployment on all nodes in the system. Foggy's architecture is based on an orchestration server responsible for monitoring the resources in the nodes and dynamically adapting the software allocation according to the rules defined by the user. However, Foggy's software allocation rules can only be configured according to fixed hardware characteristics of the nodes, i.e., node selection does not depend on dynamic system metrics such as latency, bandwidth consumption, and power consumption. These QoS factors should also be considered for software allocation decisions in fog nodes. Additionally, Foggy does not monitor the state of the running docker containers to detect and fix failures through actions such as rollback to the previous stable version or redeployment of the software container.

\subsubsection{Offloading tasks}
The processing tasks executed at the edge/fog nodes can be classified according to their importance and their response time required. While there are system tasks that do not require immediate processing, other tasks such as real-time data analysis are critical to the system and require low response latency. It is necessary to guarantee low latency for these critical tasks, but it is not trivial to achieve this when dynamic events occur in the system such as increased data flow from the physical layer. The adaptation strategy \textit{Offloading tasks} addresses this problem in the following way: to guarantee low response latency for critical processing tasks performed by the edge/fog nodes, non-critical tasks are offloaded to the cloud servers to free up capacity in the edge/fog nodes. However, it is necessary to establish when it is really necessary to offload tasks to the cloud servers. In contrast, offloading tasks from cloud servers to edge/fog nodes is also possible. This is done to improve QoS such as latency, as long as the edge/fog nodes have the necessary resources to execute the task.

S6 proposes an architecture that coordinates data processing tasks between an edge node and the cloud servers. The edge node performs data processing tasks with the data collected by devices in the physical layer of the system. A monitoring component frequently checks the CPU usage of the edge node, and every time the value exceeds a usage limit (75\%) one of the non-critical tasks executed by the node is offloaded to a cloud server. This frees up resources on the fog node for processing tasks that require low latency. However, offloading of tasks between edge/fog nodes is not considered. Before moving tasks to cloud servers, the offload tasks between neighboring edge/fog nodes that have the necessary resources available should be considered to take advantage of edge and fog computing. In particular, response latency is lower for tasks that can be executed in the edge/fog layer rather than in the cloud layer. Additionally, decisions to move tasks from one node to another node or to a cloud server could be determined by other factors such as latency, RAM usage, power consumption, and battery level (if the node is battery powered). These factors must be monitored and analyzed to make intelligent offloading decisions according to the QoS requirements of the system.

S27 proposes a fog computing framework for cognitive portable ground penetrating radars (GPRs). An offloading policy decides where to execute the sensor data analysis tasks (in mobile nodes of the physical layer or in fog nodes). The offloading policy is based on two metrics: (1) power limitations for mobile nodes that are battery powered, and (2) the efficiency of the node to perform the task. However, cloud servers are not evaluated by the offloading policy.

\subsection{Relationship between events and adaptations}
\label{sub:rel-events-adaptations}

Table \ref{table:events-strategies} presents the relationship between the dynamic events and the adaptation strategies to address them. 
The most used strategies are \textit{Data flow reconfiguration} (A1) and \textit{Offloading tasks} (A4). Strategy A1 is used to ensure system QoS for five dynamic events: (1) \textit{Client mobility} because data traffic from devices joining the system can be routed to assign services to attend them; (2) \textit{Dynamic data transfer rate} because it is possible to dynamically balance the variable data flows and distribute processing workloads among the edge/fog nodes; (3) the adaptation for \textit{Important event detected by sensors} depends on domain requirements, for example, it is possible to redirect sensor data to the nearest nodes to reduce latency (as in S8); (4) \textit{Network connectivity} because it is possible to dynamically control data flow to ensure QoS for variations in network connectivity; and (5) \textit{Failures and software aging} because it is possible to reroute data traffic when failures are detected on a node. 

The adaptation strategy \textit{Auto Scaling of services and applications} (A2) enables automatic adjustment of system capacity to ensure performance. This strategy enables the system to support variations in the workloads of edge/fog nodes and cloud servers. Therefore, the A2 strategy is used to guarantee system QoS for dynamic events that involve an increase in data processing at nodes such as \textit{Client mobility}, \textit{Dynamic data transfer rate}, and \textit{Important event detected by sensors}.

The strategy \textit{Offloading tasks} (A4) is used to guarantee QoS at least in the critical tasks of the system. This strategy frees up processing resources on the edge/fog nodes to address critical tasks. Meanwhile, low critical tasks can be offloaded to neighboring edge/fog nodes or to cloud servers. In the literature, the A4 strategy is used to ensure the QoS of critical tasks in dynamic events such as \textit{Client mobility}, \textit{Dynamic data transfer rate}, \textit{Important event detected by sensors}, and \textit{Network connectivity}.

Finally, the strategy of adaptation \textit{Software deployment and upgrade} is only used by the studies that deal with the event \textit{Failures and software aging}. The deployment of updates or new software versions in the edge/fog nodes allows for the repair of problems detected in the software. Operations such as rollback, whose function is to release the previous stable version, can also alleviate problems due to software failures in the upgrade process.

\begin{table*}[ht]
  \caption{Connection between events and strategies}
  \label{table:events-strategies}
  \begin{tabular}{l l l l l l}
  \hline
  Dynamic event  & \hl{Data flow} & \hl{Auto scaling of services} & \hl{Software deployment} & \hl{Offloading} \\
  & \hl{reconfiguration} & \hl{and applications} & \hl{and upgrade} & \hl{tasks} \\
  \hline
   Client mobility & X & X & & X \\
   Dynamic data transfer rate & X & X & & X \\
   Important event detected by sensors & X & X & & X \\
   Network connectivity & & & & X \\
   Failures and software aging & X & & X & \\
  \hline
  \end{tabular}
\end{table*}


\section{Threats to validity}
\label{Sec:threats-validity}

According to Kitchenham and Brereton \cite{kitchenham2013systematic}, there is no way to completely avoid personal bias in a literature review. Although some stages of the SLR were developed by a single researcher, we looked for methods to ensure the quality of the results. These methods are described below.

The main threats come from the screening and data collection stages. In particular, the first and second screening of inclusion and exclusion criteria are to a certain extent subjective processes that may lead to misclassification of studies. The process of filtering the studies was carried out by the first author of this paper. To ensure the quality of the filtering process, another expert researcher in the field (not a co-author of this study) carried out the same process with the same inclusion and exclusion criteria. This researcher took 25\% of the initial studies and applied the inclusion and exclusion criteria. The agreement of results between the first author and the external researcher was 85\% of the studies analyzed; the small disagreement difference is due in part to the fact that the external researcher discarded studies that addressed overly specific problems of a domain. We had a face-to-face meeting with the external researcher to understand the differences in the classification of 15\% of the papers and make final decisions. 

Another threat to validity is the subjectivity in extracting and analyzing information from studies (Sections \ref{sec:data-collection} and \ref{sec:data-analysis}), which can result in misunderstandings. One researcher extracted the data from the studies, and the rest of the authors of this SLR checked the extraction. We scheduled regular meetings to analyze the information; particularly, the data that answered the research questions. In these meetings we discussed disagreements and when necessary, we studied the text and diagrams of the conflicting study.

The list of events and adaptations we suggest is related to another threat to validity. As we did not have a unified preliminary list of events and adaptations, we defined them based on the findings of the studies. To mitigate the risk of omitting any event or adaptation, we apply an iterative content analysis method to continuously evolve the list of events and adaptations found in the studies. All the authors met together to discuss the studies that contributed a new event or adaptation to our classification. In this way we ensure that the events and adaptations addressed by the SLR studies conform to the proposed categorization.


\section{Summary and future directions}
\label{Sec:summary-future-directions}

\hl{In this section, we synthesize the findings and then elaborate on research topic that deserve further exploration.}

\subsection{Summary of the results}
\label{subsec:results}
\hl{Beyond the detailed analysis provided in the previous section we would also like to outline some overall conclusions.} 

\begin{itemize}
    \item To detect dynamic events during system operation, different QoS metrics and resource consumption are constantly monitored. The monitoring mechanisms proposed by the literature mainly focus on measuring resource consumption (CPU, memory, and storage) \cite{villari2016osmotic}. However, other factors should also be monitored and taken into account \hl{to accurately detect dynamic events}. For example, power consumption is an important metric for edge nodes that are powered by batteries.

    \item The monitored QoS metrics could be analyzed at a later stage to make system improvement decisions. This is one of the most important stages (known as Feedback) suggested by the DevOps practices \cite{bass2015devops} \cite{ebert2016devops}. Surprisingly, none of the studies analyzed in this review propose a method for recording and storing the data in a storage system so that they can be consulted later. 

    \item One of the technologies that is gaining strength for software deployment is containerization \cite{brogi2017container} \cite{von2019lightweight}. Most studies use docker containers to package and deploy software on the edge/fog nodes \cite{pahl2015containers}. One challenge for orchestrating containers in IoT systems is related to efficient allocation decisions between the edge and fog layer nodes \cite{santos2019resource}. The selection of the nodes to deploy the containers can impact the system on factors such as performance and availability.

    \item Finally, to control the deployment of new software versions decreasing the risk of process failures and increasing reliability, software deployment patterns (such as rolling deployment, blue-green, and canary) has been proposed \cite{ahmadighohandizi2016application}. While these patterns have been widely used in cloud-based applications, in IoT systems with distributed architectures it have been seldom explored.
\end{itemize}

\subsection{Future directions}
\label{subsec:future-directions}

The design of IoT systems involves coping with several challenges to ensure a good QoS even when considering the dynamic nature of the IoT environment. Some specific challenges were pointed out by the studies analyzed in this paper. Indeed, the conclusions above suggest already some areas that are not yet fully developed even if some works start to appear that address them.

Nevertheless, in this section, we want to highlight additional significant open challenges we believe need to be addressed to improve current adaptation strategies.

We classified the issues and challenges as follows.

\subsubsection{Monitoring and logging the dynamic events themselves}

Monitoring the system infrastructure is a key process in the design of a self-adaptive architecture. However, designing a continuous, scalable, resilient, and non-intrusive monitoring system for IoT systems is a challenge. In the literature, efforts are focused on designing strategies to adapt the IoT system at run time. But self-adaptations for system monitoring components also require attention. For example, according to the state of the infrastructure, the monitoring system must self-adapt to the characteristics of the heterogeneity of devices (e.g. gateways, servers, switches, and user devices), heterogeneity according to virtualization (e.g. virtual machines, containers, and pods), and scalability (join and leave of devices).

In addition to detecting events at runtime, monitoring data is also important for analyzing historical data and making decisions to improve its architecture. One of the critical stages of DevOps is monitoring and feedback. This stage involves constantly monitoring the system, storing data, and then analyzing data to provide feedback to developers. Through the analysis of the data, improvement decisions are made for the next DevOps iteration. However, it is necessary to effectively monitor and store the data for historical queries and analysis to identify system improvements. Logging of monitoring data implies the design of a domain model that abstracts the main concepts of self-adaptive and distributed IoT architectures. For example, concepts such as the different types of devices (including their resources, location, and hardware characteristics), QoS metrics, dynamic events, and adaptations. In addition, to persist time series, events, and metrics, it is necessary to select an appropriate scalable storage system such as InfluxDB\footnote{https://www.influxdata.com} or TimescaleDB\footnote{https://www.timescale.com}.

\subsubsection{Software deployment on heterogeneous devices}

Some adaptation strategies such as service auto-scaling, software deployment, and upgrades involve the deployment of new software versions in the different layers of the system architecture. Container-based virtualization and hypervisor-based virtualization are widely used in the edge/fog and cloud layers for software deployment. Container-based virtualization improves performance and efficiency when compared to hypervisor-based virtualization \cite{singh2016containers}. While a virtual machine requires a complete installation of the operating system, a software container shares the host OS kernel, binaries, and libraries. Additionally, containers are faster to create or migrate when compared to virtual machines. Therefore, containers are preferred at the edge/fog layer of the system, and there are tools such as Kubernetes\footnote{\url{https://kubernetes.io}}, K3S\footnote{\url{https://k3s.io}}, and Docker Swarm\footnote{\url{https://docs.docker.com/engine/swarm/}} to manage and orchestrate software deployments in containers. However, there are challenges for software deployment and migration in the edge/fog layer nodes of IoT systems. One of the present challenges is related to making intelligent allocation decisions to guarantee QoS. When deploying or moving an application in the system, it is necessary to select the edge/fog nodes that have enough resources for the operation of the application, and to offer the appropriate QoS. Tools like Kubernetes provide functionality through the scheduler component to select the appropriate node that will host the container with the new software version, but this component only checks the resources requested (CPU and RAM) by the container. Other factors such as energy consumption, network latency, reliability, and bandwidth usage should be considered when making allocation decisions.

Another challenge relates to the managing of software deployments on physical layer devices. The heterogeneity of sensor and actuator devices makes it difficult to deploy software at this layer due to the variety of communication protocols and software languages supported. \hl{Some physical layer devices do not support remote upgrades, nor virtualization, or containerization.}

\subsubsection{Machine learning for self-adaptable systems}
\label{subsection:machine-learning}

Machine learning systems can automatically identify normal and abnormal patterns and alert a client or third parties when things deviate from observed standards, without requiring prior configuration by human operators. For IoT systems, learning algorithms can also help to prevent disruptive events affecting system availability and QoS. For example, to predict when a hardware component might fail to take action and avoid system downtime. Predictive maintenance of physical devices is one of the tasks that can be efficiently forecasted by algorithms learning. For example, manufacturing systems require maintenance of their machinery to avoid system failures and downtimes. While there are traditional challenges for the design of a learning algorithm such as the selection of the efficient model, the amount of data, and data cleaning, there are also other problems related to the technologies and processes to obtain the data or features. For example, the monitoring of non-functional properties such as accuracy, frequency, sensitivity, and drift is one of the challenges due to the heterogeneity of IoT devices in the physical layer.

\subsubsection{Global self-adaptive architecture}

The studies included in this SLR propose techniques and strategies to address at most two of the dynamic events. However, in some scenarios or domains, it is necessary to propose solutions to support various/simultaneous dynamic events. For example, in the scenario illustrated in Section \ref{sec:intro}, the IoT ventilation monitoring and control system for underground mines can present several types of dynamic events. Another example is a smart city system, which synchronizes the basic functions of a city based on seven key components, including natural resources and energy, transport and mobility, buildings, life, government, economy, and people \cite{Colistra2019}. Due to the large number of IoT devices considered, a smart city system can experience all the dynamic events that we have identified in Table \ref{table:dynamic-events}. 

Therefore, it is necessary to design a general architecture for IoT systems with components to monitor, detect events, and self-adapt the system: an architecture with the ability to adapt to various dynamic events. \hl{For example, a system that can detect failures in software updates and perform operations such as software rollback, while supporting new devices being added to the system. This same system could also support other types of events such as dynamic data transfer rate and network connectivity failures.}

For designing this general self-adaptive architecture for IoT systems, some base technologies are especially promising. For example, the MQTT communication protocol is ideal for IoT applications since it presents advantages concerning scalability, asynchronism, decoupling between clients, low bandwidth, and power consumption. Regarding virtualization technology, containerization offers several advantages for software deployment in IoT systems. In particular, it is possible to deploy containers on various types of hardware and operating systems, something very useful considering the heterogeneity of nodes in the edge/fog layer. For example, it is possible to deploy a container with an application on both a RaspberryPI\footnote{\url{https://www.raspberrypi.org}} and a Linux server.


\section{Related Work}
\label{Sec:rel-work}

Literature reviews have been presented in the IoT field, mainly focused on different specific sub-domains such as smart cities, cyber-physical systems, or industrial IoT. For example, \cite{liao2018industrial} presents a review of standards, technologies, and implementations of industrial IoT applications, \cite{alaa2017review} provides a review of the use of smart home applications and open challenges, and \cite{alavi2018internet} elaborates on key features and applications of the IoT paradigm to support sustainable development of smart cities. These studies focus on describing the current state of IoT applications for a specific domain (industrial IoT, smart homes, and smart cities respectively), and identifying challenges and opportunities for future research in the area.

With respect to literature reviews more closely related to our research we found three studies, each of them is discussed in the following paragraphs and compared with our work.

Mekuria et al. \cite{mekuria2019smart} contribute a SLR about smart home reasoning systems (SHRS). The reasoning techniques for these systems are characterized and analyzed to discuss the challenges in smart living environments. The requirements, assumptions, strengths, and limitations of SHRSs are discussed. Although SHRSs involve control and adaptations operations, the aim of this study is not to classify system architecture adaptation strategies and dynamic events that impact QoS.

Muccini et al. \cite{muccini2016self} present an SLR to analyze the adaptation strategies at the level of cyber-physical systems architecture. From the findings, future work opportunities and challenges are identified. The target topic of \cite{muccini2016self} is complementary to our SLR. However, our study deals with both cyber-physical systems and IoT systems. In addition, we classify and analyze the dynamic events that cause adaptations due to the changing environment. 

Nguyen et al. \cite{nguyen2019advances} depict a systematic review to identify and analyze the most significant approaches in deployment and orchestration for IoT. This SLR focuses on describing the technical details of the orchestration, reliability aspects (including monitoring, adaptation, and shared access to resources), and the technical challenges for future research. However, the adaptations analyzed are limited to solutions that consider a software orchestrator. In addition, the dynamic events that induce the adaptations are not analyzed.

In a nutshell, to the best of our knowledge, there are no literature reviews that comprehensively target self-adaptation strategies, dynamic events and their impact in the quality attributes of IoT systems. Therefore, the added value of this SLR with respect to the state-of-the-art is to present an overview of self-adaptive IoT systems, the strategies used within, the events that induce the adaptations, and some gaps that need to be addressed.


\section{Conclusion}
\label{Sec:Conclusion}

We have conducted a systematic literature review to study the dynamic events that impact the QoS of IoT systems, to analyze the strategies implemented by the literature to address them, and to identify the weaknesses of the approaches found in the state-of-the-art.

We identified six types of dynamic events and four adaptation strategies in response to the events. \textit{Dynamic data transfer rate} is the event most addressed by the literature, while \textit{Data flow reconfiguration} is the strategy most used in the studies. 

Monitoring the resource consumption of the edge/fog nodes is one of the most used strategies to detect some dynamic events of the system. In particular, the consumption of CPU and RAM memory are metrics that are frequently monitored to identify events such as \textit{Dynamic data transfer} and \textit{Failures and software aging}. Other factors such as power consumption, availability, latency, throughput, and bandwidth are poorly monitored, but these factors should be considered to accurately detect dynamic events.

As further work, we plan to advance in the opportunities discussed in the previous section. We aim to address the challenges and open gaps related to the software deployment on heterogeneous devices.


\begin{backmatter}

\section*{Acknowledgements}
We would like to thank researcher Wilmer Rubio for contributing in one of the tasks to address threats to validity: to conduct the process of filtering studies independent of our process, to compare them and avoid misclassification.

\section*{Funding}
This work has been partially funded by the Spanish government (LOCOSS project - PID2020-114615RB-I00), the Colombian government (\textit{Becas del Bicentenario} program - Art. 45, law 1942 of 2018), and the ECSEL Joint Undertaking (JU) under grant agreement  No 101007260. The JU receives support from the European Union’s Horizon 2020 research and innovation programme and Netherlands, Finland, Germany, Poland, Austria, Spain, Belgium, Denmark, Norway.

\section*{Abbreviations}
SLR: Systematic literature review; IoT: Internet of things; QoS: Quality of service; SDN: Software-defined networking; MQTT: Message queue telemetry transport; GPR: Ground penetrating radar; SHRS: Smart home reasoning system.

\section*{Availability of data and materials}
Please contact authors for data requests.


\section*{Competing interests}
The authors declare that they have no competing interests.


\section*{Authors' contributions}
IA contributed to the literature search process, application of inclusion and exclusion criteria, study quality assessment, data collection and analysis. KG, HC, and JC collaborated in the formulation of the problem and research questions, the definition of inclusion and exclusion criteria, and participated in the extraction and analysis of data from the studies. All authors write, read, and approved the final manuscript.



\bibliographystyle{vancouver} 
\bibliography{references}      


\end{backmatter}
\end{document}